\documentclass[
reprint,
amsmath,amssymb,
aps,
pra,
longbibliography,
floatfix
]{revtex4-2}
\usepackage{graphicx}
\usepackage{color}
\usepackage[utf8]{inputenc}
\usepackage{dcolumn}
\usepackage{bm}
\usepackage{hyperref}
\usepackage[capitalise]{cleveref}
\usepackage[mathlines]{lineno}
\usepackage{multirow}
\usepackage{braket}

\newcommand{\ppt}{\frac{\partial}{\partial t}}

\newcommand{\Av}{{\bf A}}

\newcommand{\Ev}{{\bf E}}
\newcommand{\As}{{A_0}}
\newcommand{\p}{{\bf p}}

\newcommand{\Op}{\mathcal{O}}
\newcommand{\rv}{{\bf r}}
\newcommand{\vv}{{\bf v}}

\newcommand{\me}{{m}}
\newcommand{\TAS}{\Delta \mathcal{E}}  
\newcommand{\FTTAS}{\Delta \tilde{\mathcal{E}}}
\newcommand{\DTAS}{\Delta \dot{\mathcal{E}}}

\def\bf#1{\mathrm{\mathbf{#1}}}

\def\ket#1{| #1 \rangle}
\def\bra#1{\langle #1|}

\newcommand{\Ej}{\epsilon_j}
\newcommand{\Ejp}{\epsilon_{j'}}
\newcommand{\Ef}{\epsilon_f}
\newcommand{\Efj}{\epsilon_{fj}}
\newcommand{\Efjp}{\epsilon_{fj'}}
\newcommand{\Ejjp}{\epsilon_{jj'}}
\newcommand{\TME}{\bra{j'}\hat{p}_z\ket{f}\bra{f}\hat{p}_{z}\ket{j}}
\newcommand{\TMEb}{\bra{j'}\hat{p}_z\ket{n}\bra{n}\hat{p}_{z}\ket{j}}
\newcommand{\TMEc}{\bra{j'}\hat{p}_z\ket{\epsilon_k}\bra{\epsilon_k}\hat{p}_{z}\ket{j}}

\newcommand{\Enjp}{\epsilon_{nj'}}

\newcommand{\Ekjp}{\epsilon_{kj'}}
\usepackage{MnSymbol}
\usepackage{physics}

\begin{document}

\preprint{APS/123-QED}

\title{Gauge-invariant absorption of light from a coherent superposition of states }
\author{Axel Stenquist\,
}
\author{Felipe Zapata\,
}
\email{felipe.zapata@matfys.lth.se}
\author{Jan Marcus Dahlström\,
}
\email{marcus.dahlstrom@matfys.lth.se}

\affiliation{Department of Physics, Lund University, 22100 Lund, Sweden.}

\begin{abstract}
\noindent 
Absorption and emission of light is studied theoretically for excited atoms in coherent superposition of states subjected to isolated attosecond pulses in the extreme ultraviolet range. A gauge invariant formulation of transient absorption theory is motivated using the energy operator from Yang's gauge theory. 
The interaction, which simultaneously couples both bound and continuum states, is simulated by solving the time dependent Schrödinger equation for hydrogen and neon atoms. A strong dependence on the angular momentum and the relative phase of the states in the superposition is observed. Perturbation theory is used to disentangle the fundamental absorption processes and a rule is established to interpret the complex absorption behaviour. It is found that non-resonant transitions are the source of asymmetry in energy and phase, while resonant transitions to the continuum contribute symmetrically to absorption of light from coherent superpositions of states. 
\end{abstract}
  
\maketitle

\section{Introduction}\label{introduction}

Pulses of attosecond temporal duration in the extreme ultraviolet (XUV) regime can be created through a nonlinear optical process that is known as high-order harmonic generation (HHG) \cite{antoine_attosecond_1996}. 
As a result, coherent dynamical processes in quantum systems can be studied on the attosecond timescale \cite{corkum_attosecond_2007}. 
In 2010, the first attosecond transient absorption spectroscopy (ATAS) experiment was conducted using a pump--probe setup to investigate the dynamics of ions in superposition states \cite{goulielmakis_real-time_2010}. 
Here, and in subsequent works, an intense ultra-short laser field was used to create ions by strong-field ionization, while transient absorption of a weak XUV attosecond pulse was monitored to interpret the evolution and coherence of the ions  \cite{goulielmakis_real-time_2010,wirth_synthesized_2011,sabbar_state-resolved_2017}. Modified ATAS schemes have been used to investigate various light--matter phenomena, such as Autler-Townes splittings, Lorentz-Fano line shapes and light-induced structures in photoabsorption spectra 
\cite{wang_attosecond_2010,ott_lorentz_2013,ott_reconstruction_2014,chu_absorption_2013,petersson_attosecond_2017,chew_attosecond_2018,birk_attosecond_2020,hartmann_attosecond_2019,leshchenko_kramerskronig_2023,shi_x-ray_2020}. Recent ATAS experiments have triggered the electronic exchange interaction in complex systems such as SF$_6$ molecules, giving the opportunity to laser control effective electron-electron interactions in molecular systems \cite{rupprecht_laser_2022}. 
%
ATAS has also been used to study dynamics in semiconductors, such as band-gap dynamics in silicon \cite{schultze_attosecond_2014} and separate electron and hole relaxation dynamics in germanium \cite{zurch_direct_2017}.   
Reviews on experimental developments of ATAS are given in Refs.~\cite{beck_probing_2015,geneaux_transient_2019}. 

The foundation of our theoretical work is based on the ATAS tutorial by Wu \textit{et al.} \cite{wu_theory_2016}. 
In the present article, however, the ATAS theory is reformulated to be consistent with Yang's gauge theory, which is based on the 
so-called {\it energy operator} \cite{yang_gauge_1976}, and it forms a natural finale to the question of optimal gauge in time-resolved ATAS simulations \cite{dahlstrom_attosecond_2017,zapata_implementation_2021}.
Previous characterizations of ATAS structures have mainly been focused on bound states driven or dressed by a laser field using few-state models, see \textit{e.g.} Refs.~\cite{wu_theory_2016,rorstad_analytic_2017,leshchenko_kramerskronig_2023}. 
ATAS features in this regime were characterized in Ref. \cite{rorstad_analytic_2017} through the creation of an analytical model. 
Macroscopic propagation of the fields can give rise to additional effects, but such effects can be avoided by considering thin optical media \cite{wu_theory_2016}. Coherent superpositions of $s$-wave Rydberg states in hydrogen and helium atoms have been predicted to show dependence on the relative phase of the states in the superposition when the XUV field couples to the $p$-wave continuum \cite{dahlstrom_attosecond_2017}, but experimental studies of continuum effects using ATAS are rare \cite{birk_attosecond_2020}. Analogously, coherent superpositions have been investigated in ionization studies, where a clear dependence on the relative phase is found \cite{fechner_strong-field_2014,pabst_eliminating_2016,milosevic_high-order_2022,klunder_reconstruction_2013}. This type of phenomena is often referred to as quantum beating of superposition states.    
Although numerous investigations have been performed on a large range of complex systems, the fundamental processes that lay the foundations of ATAS have not yet been systematically explored from coherent superpositions of Rydberg states in atoms. Here a perturbation theory model is presented, which goes beyond the use of few-state models, to disentangle the key processes in atomic ATAS. In this way, it is possible to identify two kinds of light-matter interaction processes: {\it resonant} and {\it off-resonant} processes, by their different symmetries in ATAS experiments. Results for hydrogen and neon are presented, but the general conclusions are expected to be valid in any atom excited into a coherent superposition of Rydberg states. 
The article is organized as follows. Section II presents the formulation of the gauge-invariant transient absorption theory.   
In section III, the disentanglement of the fundamental ATAS processes is performed. 
In section IV, results for hydrogen and neon are discussed. Finally, section V contains our conclusion and outlook. Atomic units are used throughout this text, $e=\hbar=m=4\pi\epsilon_0=1$, unless otherwise stated.

\section{Transient absorption theory}\label{theory} 
The Hamiltonian for an electron with mass $\me=1$, charge $q=-1$ and canonical momentum $\p=-i \nabla$ in presence of a classical time-dependent electromagnetic field and a static potential can be written in the {\it minimal coupling} form: 
\begin{equation}
\label{H(A,A0)}
    H(\Av,\As)=\frac{1}{2m}\left(\p-q\Av\right)^2+q\As+V,
\end{equation}
where  $\Av(\rv,t)$ and $\As(\rv,t)$ correspond to the vector and scalar potentials of the time-dependent field, and $V(\rv)$ to the static and conservative potential of the target \cite{sakurai_modern_2017}. In this work we will consider the specific case of the hydrogen atom, where the conservative potential corresponds to the Coulomb interaction between the electron and its nucleus, $V(r)=q/r$. The corresponding time-dependent Schrödinger equation (TDSE) is given by 
\begin{equation}
    \begin{split}
\label{TDSE}
i\ppt\psi(t)
&= H(\Av,\As)\psi(t)\\
&= \frac{1}{2m}\left(\p-q\Av\right)^2\psi(t)+q\As\psi(t)+V\psi(t).
    \end{split}
\end{equation}
An important property of Eq.~(\ref{TDSE}) is that it is {\it form invariant} when the wave function $\psi\rightarrow \psi'$ and the potentials $(\Av,\As)\rightarrow (\Av',\As')$ are gauge transformed in the correct way~\cite{kobe_gauge_1978}. The expectation values of physical observables must be independent of the choice of gauge,\textit{ i.e.} 
\begin{equation}
\bra{\psi(t)}\Op\ket{\psi(t)}=\bra{\psi'(t)}\Op'\ket{\psi'(t)}.
\end{equation}
  
According to the theory of Wu \textit{et al.}, the physical observables in ATAS can be derived by an energy-conservation principle, between the atom (quantum system) and the electromagnetic radiation (classical field). The exchanged energy is defined as:  
\begin{equation}
    \TAS=\int_{-\infty}^{+\infty}\DTAS(t)dt,
\end{equation}
with $\DTAS(t)$ being the instantaneous power transferred to the quantum system due to its coupling with the classical field~\cite{wu_theory_2016}. Intuitively, this instantaneous power should describe instantaneous absorption/emission processes of radiation by the atom. According to Wu \textit{et al.}, $\DTAS(t)$ should be defined as the time-derivative of the expectation value of the atomic Hamiltonian, \textit{i.e.}
\begin{equation}
\DTAS=\frac{d}{dt}\bra{\psi(t)}H(\Av,\As)\ket{\psi(t)}, 
\end{equation}
where further the particular case of the length gauge was employed. In previous works, we have stressed that $\DTAS(t)$ is an elusive quantity that is not gauge invariant ~\cite{dahlstrom_pulse_2019,zapata_implementation_2021}. The origin of this gauge ambiguity is related to the fact that the expectation value of the minimal-coupling Hamiltonian $H(\Av,\As)$ is not gauge invariant in presence of an electromagnetic field~\cite{kobe_gauge_1978}, \textit{i.e.}
\begin{equation}
\bra{\psi(t)}H(\Av,\As)\ket{\psi(t)}\neq\bra{\psi'(t)}H(\Av',\As')\ket{\psi'(t)}.
\end{equation}
%
In order to fully avoid gauge ambiguities, a consistent gauge-invariant formulation of  transient absorption theory must be used. From the best of our knowledge, such a gauge-invariant formulation has not yet been proposed in the context of ATAS. 

\subsection{Gauge-invariant formulation of ATAS} \label{theory A}
As the minimal-coupling Hamiltonian $H(\Av,\As)$ is not gauge invariant, it cannot represent a physical observable~\cite{kobe_gauge_1978}. This issue forces us to search for an unambiguous quantum-mechanical operator that can be used to describe the instantaneous energy of the quantum system. In the context of the semi-classical light-matter interaction theory, Yang~\cite{yang_gauge_1976} proposed a gauge-invariant formalism based on the so-called {\it energy operator}: 
 \begin{equation}
    \label{H(Av,0)}
        H(\Av,0)=\frac{1}{2m}\left(\p-q\Av\right)^2+V,
    \end{equation}
which satisfies the gauge condition:  
\begin{equation}
\bra{\psi(t)}H(\Av,0)\ket{\psi(t)}=\bra{\psi'(t)}H(\Av',0)\ket{\psi'(t)}.
\end{equation}
In order to associate this Hamiltonian to the instantaneous energy operator, Yang~\cite{yang_gauge_1976} applied the correspondence principle of quantum mechanics: 

\vspace{0.2cm}
\textit{``[...] an operator represents a physical quantity with a classical analogue only if the equation of motion for the expectation value of the operator is of the same form as the equation of motion for the corresponding classical Newtonian quantity''}. 
\vspace{0.2cm}

The equation of motion for the expectation value of the energy operator can be derived using Ehrenfest's theorem and is given by \cite{yang_gauge_1976,kobe_gauge_1978}
\begin{equation}
\label{Eq:EoM for energy operator}
    \frac{d}{dt}\bra{\psi(t)} H(\Av,0)\ket{\psi(t)} = \frac{q}{2}\bra{\psi(t)}\vv\cdot\Ev+\Ev\cdot\vv\ket{\psi(t)},
\end{equation}
where the velocity operator is given by
\begin{equation}
    \label{velocity_operator}
    \vv=\frac{1}{m}\left(\p-q\Av\right),
\end{equation}
and $\Ev(\rv,t)$ the electric field from the time-dependent external field (which does not include any contribution from the conservative potential, $V$).  
If a classical particle is subject to a combination of forces, $\mathbf{F}=\mathbf{F}_0+\mathbf{F}_1$, consisting of a conservative force, $\mathbf{F}_0(\rv)=-\nabla V$, and a non-conservative (in our case explicitly time-dependent) force, $\mathbf{F}_1=\mathbf{F}_1(t)$, then the change in total energy of the particle $E_T$ is given by $dE_T=\mathbf{F}_1\cdot \mathbf{v}dt$, {\it c.f.} Appendix A of Ref.~\cite{yang_gauge_1976}.   
%
Thus, the time-derivative of the energy operator, in \cref{Eq:EoM for energy operator}, is a power caused by the electric-field force, $\mathbf{F}_1=q\Ev$.  
By means of the correspondence principle, the Hamiltonian $H(\Av,0)$ can be associated with the {\it instantaneous energy operator} of the quantum system. 
Consequently, the gauge-invariant power in ATAS theory should be defined as
 \begin{eqnarray}
\DTAS(t)=\frac{q}{2}\bra{\psi(t)}\vv\cdot\Ev+\Ev\cdot\vv\ket{\psi(t)}.
\end{eqnarray}
Within the electric-dipole approximation, and assuming the Coulomb gauge, $\nabla\cdot\Av=0$, the gauge-invariant instantaneous power can be written as 
\begin{eqnarray}
\DTAS(t)&=&q\;\Ev(\mathbf{0},t)\cdot \vv(t),
\end{eqnarray}
where $\vv(t)=\bra{\psi(t)}\vv\ket{\psi(t)}$ is the gauge-invariant expectation value of the velocity operator given by Eq.~(\ref{velocity_operator}).

If a linear polarised laser field along the $z$-axis is considered, the power is given by
\begin{eqnarray}
\label{power_z_ginv}
\DTAS_z(t)&=&q\,E(t)\,v_z(t),
\end{eqnarray}
and the time-dependent exchanged energy is 
\begin{equation}
\label{eq:time_res_energy_gain}
    \TAS_z(t)=q\int_{-\infty}^{t}dt'\,E(t')v_z(t'),
\end{equation}
where $v_z(t)$ is the expectation value of the velocity operator along the polarization axis.   

In order to derive the energy-domain picture, which is required for comparison with experimental measurements \cite{wu_theory_2016}, the total exchanged energy in Eq.~(\ref{eq:time_res_energy_gain}) is rewritten
as 
\begin{equation}
    \TAS_z(\infty)=2q\int_{0}^{+\infty}\mathrm{Re}[\tilde{v}_z(\omega)\tilde{E}^*(\omega)]d\omega,
\end{equation}
where $\tilde{v}_z(\omega)=\tilde{v}_z^{*}(-\omega)$ and $\tilde{E}(\omega)=\tilde{E}^*(-\omega)$ are the Fourier transforms$\,^\text{1}$ of the real functions ${v}_z(t)$ and $E(t)$, respectively.
\begin{table}[b!]
\begin{flushleft}
  $^\text{1}$
  Fourier transform convention: $\tilde f(\omega) = \dfrac{1}{\sqrt{2\pi}} \dint_{-\infty} ^\infty f(t) e^{i\omega t} dt $ and $ f(t) = \dfrac{1}{\sqrt{2\pi}} \dint_{-\infty} ^\infty \tilde f(\omega) e^{-i\omega t} d\omega$.
\end{flushleft}
\end{table}
This implies that the energy-resolved gauge-invariant gain by the atom is given by 
\begin{equation}
\label{FTTAS_ginv}
    \FTTAS_z(\omega)=2q\,\mathrm{Re}[\tilde{v}_z(\omega)\tilde{E}^*(\omega)],
\end{equation}
where the energy argument is positive, $\omega\geq0$. 
Alternatively, by inserting the Ehrenfest relation: $\dot z(t) = v_z(t)$ into \cref{eq:time_res_energy_gain}, the gauge-invariant energy-resolved gain can be written as 
\begin{equation}
\label{Wu-huu}
    \FTTAS_z(\omega)=2q\omega\,\mathrm{Im}[\tilde{z}(\omega)\tilde{E}^*(\omega)],
\end{equation}
which is identical to the expression derived by Wu {\it et al.} in length gauge \cite{wu_theory_2016}. 
While the energy-domain expressions for absorption, Eqs.~(\ref{FTTAS_ginv})~and~(\ref{Wu-huu}), are fully consistent with previous results, {\it c.f.} Refs.~\cite{wu_theory_2016,dahlstrom_attosecond_2017,zapata_implementation_2021}, the time-dependent exchange energy in Eq.~(\ref{eq:time_res_energy_gain}) differs from the corresponding expression derived from a length-gauge Hamiltonian \cite{dahlstrom_attosecond_2017}. This ``paradox'' is now lifted, because it is easy to understand that the gauge-invariant power  can be substituted by the (incorrect) length-gauge expression: $qE(t)\dot z(t)\rightarrow -q\dot E(t) z(t)$, only under time integrals with boundary terms that vanish in partial integration. In practical situations, such conditions are met because pulses vanish at early and late times, but we believe that these insights are useful to better interpret  ATAS experiments in the time domain.  

\subsection{Implementation of the gauge-invariant theory}
In our numerical implementation we consider a linearly polarized light pulse along the  $z$-component, within the electric dipole approximation, such that 
$\Av(\rv,t)\approx \Av(\mathbf{0},t) = A(t)\hat z$. 
The {\it velocity gauge} wavefunction, $\psi^V(t)$, is obtained from Eq.~(\ref{TDSE}) on the form 
\begin{equation}
    \begin{split}\label{TDSE_VG}
i\ppt\psi^V(t)
&=H(\Av^V,\As^V)\psi^V(t) \\
&=\left[\frac{\p^2}{2m}+V-\frac{q}{m} A^V(t)p_z\right]\psi^V(t),
    \end{split}
\end{equation}
where the gauge transformations: $A^V(t)=A(t)$ and $\As^V(t)=-\frac{q}{2m}A^2(t)$ have been chosen. The electric field is related to the vector potential as 
\begin{equation}
\label{Efield_VG}
    E(t)=-\ppt A^V(t).
\end{equation}
The expectation value of the velocity operator is 
\begin{equation}
\label{velocity_VG}
    v_z(t) = \frac{1}{m}\left[p_z^V(t)-qA^V(t)\right],
\end{equation}
where $p_z^V(t)$ is the expectation value  of the $z$-component of the canonical momentum computed in velocity gauge; \textit{i.e.} $\bra{\psi^V(t)}p_z\ket{\psi^V(t)}$. 
Thus, Eqs.~(\ref{Efield_VG})~and~(\ref{velocity_VG}) can be used to rewrite Eq.~(\ref{FTTAS_ginv}) as follows, 
\begin{equation}
\begin{split}
\label{FTTAS_ginv_VG}
    \FTTAS_z(\omega)
    &= \frac{2q}{m}\omega\;\mathrm{Im}[\tilde p_z^V(\omega)\tilde{A}^{V*}(\omega)],
\end{split}
\end{equation}
where the second term, proportional to $\tilde A^V(\omega) \tilde{A}^{V*}(\omega)$, has been disregarded as it is real \cite{dahlstrom_attosecond_2017}. In the following, the superscript $V$ is dropped: $A^V\rightarrow A$ because all calculations will be performed in velocity gauge.  
The attosecond XUV pulse is described by a vector potential with a Gaussian-shaped envelope, defined as
\begin{equation}
    \label{Eq: vector potential}
    A(t) = A_0 \cos{(\omega_0 t + \phi)} e^{-at^2},
\end{equation}
where  $a = 2 \ln{2}/\tau_e^2$. $A_0$, $\omega_0$, $\phi$ and $\tau_e$ are the amplitude, central frequency, carrier-envelope phase (CEP) and pulse duration, respectively. 
The frequency-dependent vector potential is obtained through the Fourier transform as
\begin{equation}
    \begin{split}
\tilde{A}^\pm(\omega) 
        &= \frac{A_0}{2\sqrt{2a}} 
        \exp\left( \pm i\phi - \frac{(\omega \pm \omega_0)^2}{4a}\right),
    \end{split}
\end{equation}
where $A^\pm$ denotes $A = A^+ + A^-$. Note that in this expression the positive component is negligible for positive frequencies. 

\section{Disentanglement of fundamental processes}
\begin{figure}
\begin{center}
\includegraphics[width=0.4\textwidth]{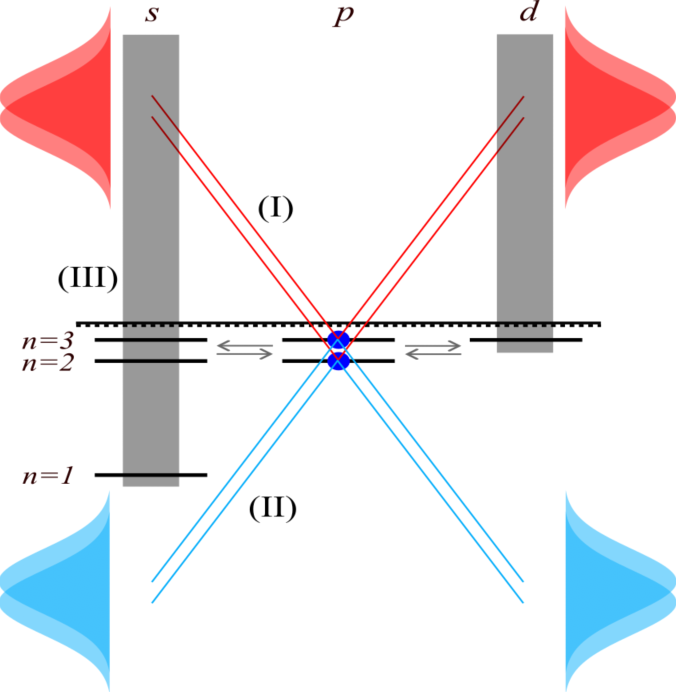}
\caption{Representation of a hydrogen-like atom prepared in a coherent superposition of states: $2p+3p$, with all processes that are induced by an attosecond XUV pulse indicated by Roman numerals. (I) Resonant transitions to the continuum (red lines). (II) Resonant transitions to the bound states, which may reside below or above the initial states in the superposition (blue lines). (III)  Off-resonant transitions to all states allowed by dipole-selection rules (grey rectangles). The bandwidth of the attosecond pulse is assumed to be larger than the separation of the non-degenerate states in the superposition, which implies that quantum beat phenomena may occur that depend on the phases (or more generally the coherence) of the states.} 
\label{fig: intro schematic}
\end{center}
\end{figure}

Consider a superposition on the general form
\begin{equation}
    \label{Eq: initial state}
    \ket{\psi_0(t)} = U_0(t,-\infty)\ket{\psi_0(-\infty)} = \sum_j^N c_j e^{-i \epsilon_j t} \ket{j}, %
\end{equation}
where $c_j$ and $\epsilon_j$ are interaction-picture amplitude and energy of stationary state $\ket{j}$, respectively, and $N$ is the total number of coherently prepared states. \cref{fig: intro schematic} shows a specific scenario where a hydrogen-like atom is prepared in a superposition of the states $\ket{2p_0}$ and $\ket{3p_0}$ with common angular-momentum quantum numbers: $\ell=1$ and $m=0$ and with equal interaction amplitudes (for simplicity denoted ``$2p+3p$'' in the following). 
Through the interaction with an attosecond XUV pulse, three fundamental processes may take place. The first process (I) is the resonant continuum contribution, represented by red lines, which consists of absorption of light due to transitions from the initial superposition to the continuum. Interference may occur as different paths to the same continuum state are allowed (absorption profiles are represented by partially overlapping Gaussian functions). The second process (II) is the resonant bound contribution, represented by blue lines, given by the emission (and absorption) of resonant dipole-allowed transitions to the bound states. The third process (III) is the off-resonant contribution, represented by grey rectangles, and it covers the whole spectrum of bound and continuum states. All processes change the angular momentum as $\ell'= \ell \pm 1$ within the electric dipole approximation. In order to characterize and disentangle these different processes, we will control the atomic quantum phases, $\varphi_j=\mathrm{arg}(c_j)$.  


\subsection{Perturbative treatment }

The fundamental ATAS processes are ``hidden'' in \cref{FTTAS_ginv_VG}, where the Fourier transform of the momentum, $\tilde p_z(\omega)$, is the key quantity, which contains the response of the electron to the electromagnetic field, $\tilde A(\omega)$. Due to the low intensity of the pulse, perturbation theory can be applied to compute the momentum $ p_z(t)=\bra{\psi(t)}\hat p _z \ket{\psi(t)}$ and its corresponding Fourier transform $\tilde{p}_z(\omega)$.

The time-dependent wave function is given by
\begin{equation}
    \ket{\psi(t)} = U(t,-\infty) \ket{\psi_0(-\infty)}, 
\end{equation}
where $U$ is the evolution operator and the initial state superposition is given by \cref{Eq: initial state}.  
The propagation is re-written in terms of the unperturbed evolution operator, $U_0$, using the well-known Dyson series expansion \cite{sakurai_modern_2017}
\begin{equation} \label{Eq: Dyson}
\begin{split}
    &\ket{\psi(t)} 
    \approx  U_0(t,-\infty) \ket{\psi_0(-\infty)} \\ 
    & \,\,\,\,\,
    - i \int_{-\infty}^t dt' U_0(t,t') H_\mathrm{int}(t') U_0 (t',-\infty) \ket{\psi_0 (-\infty)} \\
    &=  \sum_j^N c_j \ket{j} e^{-i \Ej t}  
    - i \sum_j^N \sumint_f c_j e^{-i \Ef t} \ket{f} \bra{f}\hat{p}_{z}\ket{j}    
    \\
    & \,\,\,\,\,
    \times \int_{-\infty}^t dt'A(t')  e^{i \Efj t'},
\end{split}
\end{equation}
where $\Efj = \epsilon_f - \epsilon_j$ is the difference of the energies of the states $\ket{f}$ and $\ket{j}$, the velocity-gauge interaction Hamiltonian is defined as $H_{\text{int}} = A(t)\hat p_z$ from Eq. \ref{TDSE_VG}, with $q=-1$.
\cref{Eq: vector potential} is inserted into \cref{Eq: Dyson} and the time integral is evaluated using properties of the error function. Thus, the vector potential contribution is given by
\begin{equation}
    \begin{split}
    A^\pm = &
    \int_{-\infty}^t dt' A_0\frac{e^{\pm i(\omega_0 t' + \phi)}}{2}  e^{-at^2}  e^{i \Efj t'} 
    \\
    =     
    &\frac{A_0}{4} \sqrt{\frac{\pi}{a}} e^{\pm i\phi} \exp[-\frac{(\Efj \pm \omega_0)^2}{4a}] 
    \\
    &\times \left\{ \erf\left[\sqrt{a}t-\frac{i (\Efj \pm \omega_0)}{2\sqrt{a}}\right] + 1 \right\}.
\end{split}
\end{equation}
The momentum expectation value $p_z(t)$ is computed with respect to the wave function given by \cref{Eq: Dyson} and is expressed as 
\begin{equation}
    \begin{split}
    \label{eq: time-dependent momentum expectation value}
        p_z^\pm(t) 
        = &
        \frac{- i A_0 \sqrt{\pi} }{4\sqrt{a}} \sumint_f \sum_{jj'}^N c_{j'}^* c_j \TME 
        \\ & \times 
        e^{-i \Efjp t} \left\{ \erf\left[\sqrt{a}t-\frac{i (\Efj \pm \omega_0)}{2\sqrt{a}}\right] + 1 \right\}
        \\ & \times 
         e^{\pm i\phi - \frac{(\Efj \pm \omega_0)^2}{4a}} + \text{c.c.}, 
    \end{split}
\end{equation}
where parity has been taking into account by setting matrix elements between states of the same parity to be zero. The sums over initial states are labelled with the indices $j$ and $j'$ in such a way that the expression has been simplified to its present form. Finally, $\tilde{p}_z(\omega)$ can be obtained by the Fourier transform of Eq.~(\ref{eq: time-dependent momentum expectation value}) and is given by
\begin{equation} \label{Eq: FT pz}
    \begin{split}
        \tilde p_z^\pm & (\omega) 
        = 
        \frac{A_0}{2\sqrt{2a}} \sumint_f \sum_{jj'}^N 
        \\
        & \Bigg\{ c_{j'}^* c_j \TME 
        e^{\pm i\phi -\frac{(\omega-\Ejjp\pm\omega_0)^2}{4a}} 
        \\
        & \times 
        \left[ \frac{1}{\omega-\Efjp} - i \pi \delta(\omega - \Efjp)\right]
        \\
        & - \left(c_{j'}^* c_j \TME\right)^*
        e^{\mp i \phi -\frac{(\omega + \Ejjp \mp \omega_0)^2}{4a}} 
        \\
        & \times 
        \left[ \frac{1}{\omega+\Efjp}  - i \pi \delta(\omega + \Efjp) \right]
        \Bigg\}, 
    \end{split}
\end{equation}
where we have used integration by parts to compute the product of a complex exponential factor and an error function, yielding the Fourier transform of a Gaussian. The boundary term vanishes as it becomes an infinitely fast oscillating function. In addition, exponential functions have been simplified, using the Dirac $\delta$ functions. The sum running over the final states, $\sumint_f$, is split into bound and continuum states, yielding the final expression
\begin{equation}
    \begin{split}
    \label{Eq: final momentum expectation value}
        \tilde p_z^{\pm}(\omega)
        = &  
        \frac{A_0}{2\sqrt{2a}}  \sum_{jj'}^N
        \bigg[ c_{j'}^* c_j
        e^{\pm i\phi -\frac{(\omega-\Ejjp\pm\omega_0)^2}{4a}} 
        \\&
        \times\left( O_{c-}^{jj'} + R_{c}^{jj'} + O_{b-}^{jj'} + R_{b-}^{jj'} \right) 
        \\
        &  
        - c_{j'} c_j^* 
        e^{\mp i \phi -\frac{(\omega + \Ejjp \mp \omega_0)^2}{4a}}
        \\&
        \times\left( O_{c+}^{jj'} + O_{b+}^{jj'} + R_{b+}^{jj'} \right)
        \bigg].
    \end{split}
\end{equation}

The resonant-continuum contribution, $R_{c}^{jj'}$, process I in \cref{fig: intro schematic}, is given by
\begin{equation}
\label{Eq:ResCont}
    R_{c}^{jj'} = - i \pi \TMEc \bigg|_{\epsilon_k=\omega + \Ejp},
\end{equation}
where $\bra{\epsilon_k}\hat{p}_{z}\ket{j}$ is the bound-continuum matrix element, given by \cref{Eq: TME} in \cref{appendixC}. The matrix elements are evaluated for the intermediate continuum state with the energy $\epsilon_k=\omega + \Ejp$. 
The resonant-bound contribution, $R_{b\pm}^{jj'}$, process II in \cref{fig: intro schematic}, is given by
\begin{equation}
\label{Eq:ResBound}
    R_{b\pm}^{jj'} = - \frac{i \pi}{\Delta\omega} \TMEb \bigg|_{n=(n_{j'}^{-2} \pm 2\omega)^{-1/2}},
\end{equation}
where $\bra{n}\hat{p}_{z}\ket{j}$ is the bound-bound matrix element, given by \cref{Eq: TME} in \cref{appendixC}. The matrix elements are evaluated for the intermediate bound state with principal quantum number $n=(n_{j'}^{-2} \pm 2\omega)^{-1/2}$. The expression is resolved on a numerical grid as described in \cref{appendixB}, $\Delta\omega$ being the distance between grid points. 
Finally, the off-resonant-continuum and the off-resonant-bound contributions, $O_{c\pm}^{jj'}$ and $O_{b\pm}^{jj'}$, respectively, process III in \cref{fig: intro schematic}, are given by 
\begin{equation}
\label{Eq:OffResCont}
    O_{c\pm}^{jj'} = \int_0^\infty d\epsilon_k  \TMEc\frac{\text{p.v.}}{\omega\pm\Ekjp}, 
\end{equation}
and 
\begin{equation}
\label{Eq:OffResBound}
        O_{b\pm}^{jj'} = \sum_n \frac{\TMEb}{\Delta\omega}
        \int_{\omega}^{\omega+\Delta\omega}d\omega'\frac{\text{p.v.}}{\omega'\pm\Enjp},
\end{equation}
where $\text{p.v.}$ denotes the principal value integral and $\epsilon_n$ denotes the energy of the bound state $\ket{n}$.


\section{Results and Discussion}\label{results} 
Results for hydrogen and neon atoms are presented and discussed in this section. The hydrogen atom has been chosen as a benchmark case for two main reasons: (i) all electronic states that are required for the perturbative model are analytically known, see \emph{e.g.} \cite{bethe_quantum_2013}, and (ii) the time-propagation of the TDSE can be considered numerically exact within the dipole approximation. 
In Sec.~\ref{sec:time-absorption}, time-resolved ATAS features are explored for the hydrogen atom. Subsequently, in Sec.~\ref{sec:validation}, the perturbative model is validated in the energy domain by direct comparison with numerical simulations from the TDSE. In Sec.~\ref{sec:disentangle}, the perturbative model is used to disentangle the fundamental processes in ATAS from a superposition of hydrogenic states. In  Sec.\ref{sec:distang}, the role of the different angular momentum channels is investigated. 
After validating our model, the dynamics of various coherent superpositions of states in neon is presented and interpreted in Sec.~\ref{Neatom}.

 
\subsection{Hydrogen atom} \label{Hatom}
Our numerical study is limited to equal population two-state superpositions with the following initial amplitudes,
\begin{equation}
    \label{Eq: two state initial wave packet amplitudes}
    c_1=\frac{1}{\sqrt{2}}  ; \qquad c_2=\frac{1}{\sqrt{2}}\exp(i\varphi), 
\end{equation}
where $\varphi$ is the relative superposition phase (RSP). While the case of $ns+n's$ superposition states have been considered previously \cite{dahlstrom_attosecond_2017}, the more general case of $n\ell+n'\ell$ has not been studied and, as we will show, there are subtle dependencies on the the angular momentum, $\ell$, in the superposition. Extension of our theory to more complex superpositions, which may include more states or different angular momentum, is straightforward and it may be the subject of further studies if experiments are performed on such targets in the future.    

\begin{figure*}
    \centering
    \includegraphics[width = 1\textwidth]{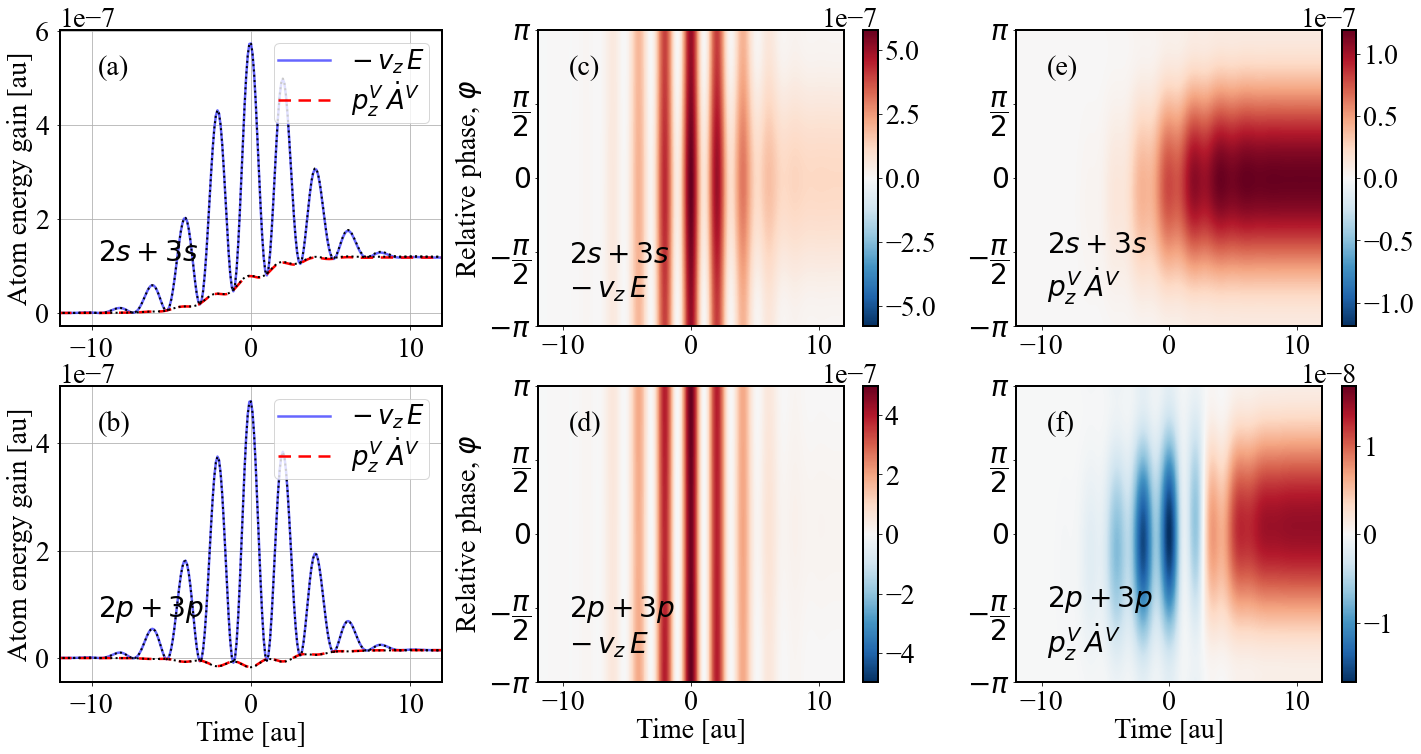}
\caption{Time-resolved energy gain of a hydrogen atom in the coherent superposition states $2s+3s$ (top row) and $2p+3p$ (bottom row) interacting with an attosecond pulse. The left column presents the total gauge-invariant gain in full lines and the relative energy gain (without the free-electron contribution) in dashed lines for the synchronized RSP, $\varphi=0$. The perturbation model results are validated by solving the TDSE numerically (dotted lines). The middle column presents the RSP-resolved total gain, whilst the right column shows the corresponding relative gain. Positive gain implies absorption (red), while negative gain implies emission (blue) of energy by the atom.}
\label{fig:time_abs}
\end{figure*}

\subsubsection{Interpretation of time-dependent absorption}
\label{sec:time-absorption}

The time-resolved energy gain of a hydrogen atom that interacts with an attosecond pulse is shown in \cref{fig:time_abs}~(a)~and~(b), for the prepared superposition states $2s+3s$ and $2p+3p$, respectively. The attosecond pulse has central frequency $\omega_0 = 1.5$ au, pulse length $\tau_e = 7$ au and vector-potential magnitude, $A_0 = 10^{-3}$ au. The energy gain is computed using \cref{eq:time_res_energy_gain} with the momentum in Eq.~(\ref{eq: time-dependent momentum expectation value}). It represents the integrated atomic power: $-E(t)v_z(t)$ in Eq.~(\ref{power_z_ginv}), which is shown in  \cref{fig:time_abs} for $2s+3s$ (c) and $2p+3p$~(d), also resolved over the RSP, $\varphi$. While the gauge-invariant energy gain is quite similar in shape and magnitude for the two cases (c-d), the final energy gain is much larger for $2s+3s$ (a) than for $2p+3p$ (b). In order to interpret this puzzling observation, we present in (a-b) a comparison of the atomic energy gain with a {\it relative} energy gain that corresponds to the gain of the atom minus the gain of a free electron with no initial velocity, $v_z(-\infty)=p_z^V=0$, as proposed in Ref.~\cite{zapata_implementation_2021}. The power of a free electron with no initial velocity is $\dot A^V(t) A^V(t)$, which obviously does not lead to any net energy gain in a laser field with $A^V(\pm\infty)=0$. The relative gain allows us to interpret the atomic absorption process with this ``virtual'' free-electron gain removed. The relative gain is computed using the following power: $p_z^V(t) \dot A^V(t)$.  Here all quantities are computed using the velocity gauge, see Eq.~(\ref{TDSE_VG}), but we stress that the results are gauge invariant due to the usage of the energy operator of Yang \cite{yang_gauge_1976}. 
%
The relative energy gain is slowly changing and we propose that this quantity can be interpreted a gradual net energy gain of the atom in the field. 
For $2s+3s$ (a) it is positive at all times, while for $2p+3p$ (b) it has an interval of energy loss during the interaction with the pulse. This energy loss is the reason for the much smaller energy gain of $2p+3p$ when compared with $2s+3s$ at the end of the pulse. The RSP-resolved energy gain in (c) and (d) are similar as they are dominated by the power of the free-electron, however significant differences between the two cases are observed in the relative energy gain presented in (e) and (f). For $2s+3s$ only absorption (positive energy gain) is observed at all RSP, while for $2p+3p$ emission (negative energy gain) is found during the beginning and middle of the pulse, while absorption is established only towards the end of the pulse. The time-resolved energy gain is symmetric around $\varphi=0$. The magnitude of the relative energy gain is small for out-of-phase RSP, $\varphi\approx \pm\pi$, while it is stronger for synchronized RSP, $\varphi\approx 0$. Finally, we have found that the CEP of the pulse, $\phi$, determines the peak structure in the energy gain, but that the CEP does not affect the total energy gain by the atom at the considered pulse parameters. 
\begin{figure*}
    \centering
    \includegraphics[width = 1\textwidth]{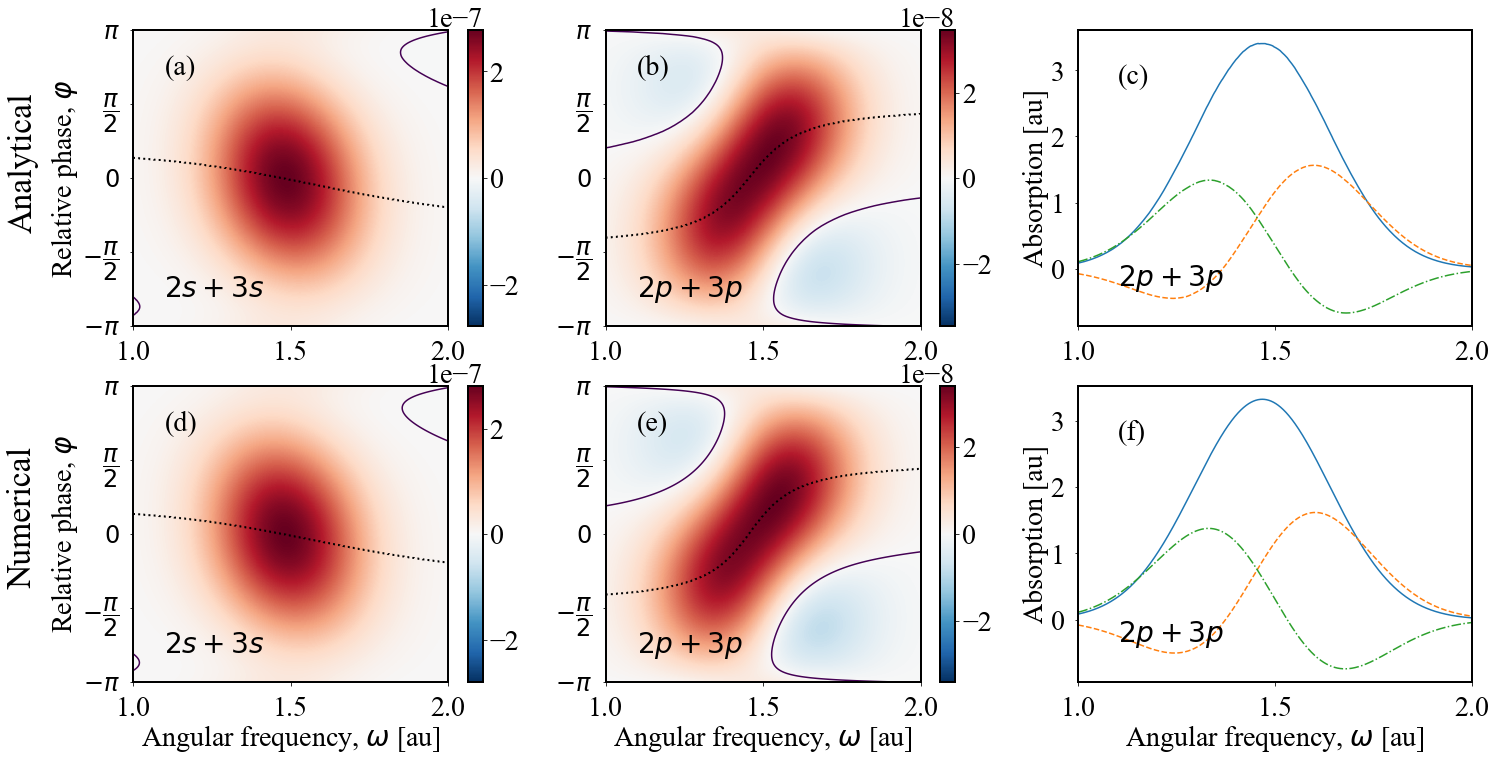}
    \caption{Analytical (top row) and numerical (bottom row) energy-resolved absorption of an attosecond pulse by a hydrogen atom in the prepared superpositions: $2s+3s$ and $2p+3p$. The data is resolved over angular frequency of the field, $\omega$, and RSP, $\varphi$. Left column presents $2s+3s$, middle column $2p+3p$, the dotted black line shows the phase of maximal absorption and the purple line is a contour showing where the absorption is zero. Negative absorption is interpreted as emission of energy to the field by the atom (blue). The right column presents $2p+3p$ with lineouts for the phases $\varphi = 0$ (solid line), $\varphi = \frac{3\pi}{4}$ (dashed line) and $\varphi = - \frac{3\pi}{4}$  (dash-dotted line).}
    \label{fig:ananum}
\end{figure*}
This is in contrast to the RSP, $\varphi$, which strongly affects the magnitude of the energy gain. 

\subsubsection{Validation of perturbation model in the energy domain}\label{sec:validation}
 
The absorption, in the energy domain, of a hydrogen atom in two different superpositions ($2s+3s$ and $2p+3p$) is computed with Eq.~(\ref{FTTAS_ginv_VG}) and shown in  \cref{fig:ananum}. In the top row, results are obtained using perturbation theory with the momentum given by Eq.~(\ref{Eq: final momentum expectation value}). In the bottom row, results are computed by numerically propagating the TDSE in the velocity gauge given in Eq.~(\ref{TDSE_VG}). 
The absorption is given as a function of the angular frequency and the phase of the superposition for an attosecond pulse with the same parameters as in \cref{sec:time-absorption}. As expected, there is good agreement between the results from perturbation theory and exact numerical propagation.  
Further, our result for the $2s+3s$ superposition is in good agreement with previous studies \cite{dahlstrom_attosecond_2017}. The maximal absorption is obtained when the two states are roughly in phase, which corresponds to the case when the atom can be photoionized without destructive quantum interference from the two states in the superposition. 
The exact phase for maximal absorption depends on the angular frequency and it is shown with a dotted black line. Interestingly, the exact phase has a negative slope over the photon energy for the $2s+3s$ case, while it has a positive slope for the $2p+3p$ case. A further significant difference between the superpositions is that the $2s+3s$ case is associated with mostly absorption of light (shown in red colour), while the $2p+3p$ superposition exhibits large spectral regions with emissions of light (shown in blue colour). \cref{fig:ananum}~(c) and (f)  show lineouts of the $2p+3p$ superposition at three different RSPs: $\varphi = 0$ and $\varphi = \pm 3\pi/4$. This demonstrates that the intricate absorption and emission phenomena are consistently manifest in both analytical and numerical results. The transition from symmetric to asymmetric curves in ~\cref{fig:ananum}~(c) and (f) is reminiscent of Fano line shapes \cite{fano_effects_1961}. While spectrally narrow atomic absorption lines have been manipulated from symmetric Lorentz line shapes to asymmetric Fano line shapes using ATAS \cite{ott_lorentz_2013}, the present result  shows that the entire broad bandwidth of attosecond pulses can be manipulated using the phases of an atom in a prepared superposition. Thus, we believe that the present result may provide a novel way to tailor the  spectral content of isolated attosecond pulses using atoms in time-dependent superpositions. 
We note that regions of emission are observed in both the energy domain and the time domain for the $2p+3p$ superposition.



\subsubsection{Fundamental processes in ATAS}
\label{sec:disentangle}

\begin{figure}
\begin{center}
\includegraphics[width=0.5\textwidth]{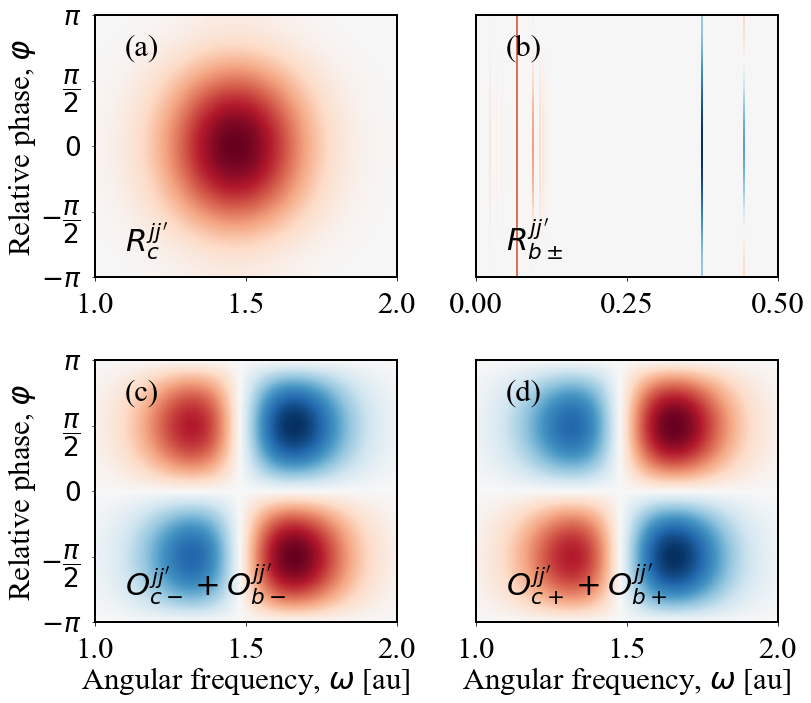}
\caption{Disentangled fundamental processes in energy-resolved absorption as a function of RSP, $\varphi$, for a hydrogen atom in superposition: $2p+3p$. (a) Resonant-continuum contribution, {\it c.f.} \cref{fig: intro schematic}\,(I). (b) Resonant-bound contribution, {\it c.f.} \cref{fig: intro schematic}\,(II). (c) and (d) off-resonant $-$ and $+$ frequency contributions, respectively, {\it c.f.} \cref{fig: intro schematic}\,(III). Transitions to bound states (below the ionization threshold) are shown in (b), while transitions to continuum states (above the ionization threshold) are shown in (a,c,d). The $R_{b-}^{jj'}$ contribution has been scaled up by a factor of 10 for clarity of view in (b).} 
\label{disentangle}
\end{center}
\end{figure}

Having verified the perturbation theory model, we now analyse its different contributions in detail.  In \cref{disentangle} we show the ATAS result for the $2p+3p$ case separated into the fundamental terms of \cref{Eq: final momentum expectation value}. These terms are illustrated in \cref{fig: intro schematic} with 
(I) being the resonant transitions to the continuum, $R_{c}^{jj'}$ in \cref{Eq:ResCont}, 
(II) the resonant transitions to the bound states, $R_{b\pm}^{jj'}$ in \cref{Eq:ResBound}, and 
(III) the off-resonant transitions, $O_{c\pm}^{jj'}$ and $ O_{b\pm}^{jj'}$, in \cref{Eq:OffResCont,Eq:OffResBound}, respectively.  
%
The resonant continuum contribution has a broad Gaussian-like shape over angular frequency, as shown in Fig.~\ref{disentangle}~(a), while the resonant bound contribution shows narrow absorption and emission lines, Fig.~\ref{disentangle}~(b). The width of the narrow lines is determined by the resolution of photon energy, see \cref{appendixB}, and the strongest absorption/emission is found for zero phase, $\varphi=0$.  
Interestingly, absorption is observed in the high-frequency emission line in the out-of-phase case, presumably due to a redistribution of energy.  
All resonant absorption features are symmetric with the phase transformation: $\varphi \rightarrow -\varphi$.

The off-resonant terms exhibit an absorption and emission checkerboard pattern. There are two off-resonant contributions, coming from the $-$ and $+$ components of $O_{c\pm}^{jj'}$ and $ O_{b\pm}^{jj'}$, as shown in Fig.~\ref{disentangle}~(c) and Fig.~\ref{disentangle}~(d) respectively. Unlike the resonant case, the absorption/emission features are antisymmetric with respect the phase transformation: $\varphi \rightarrow -\varphi$. Further, the two off-resonant contributions ($\pm$) have opposite properties. Thus, the relative magnitude of the off-resonant contributions will determine the slope of the phase for maximal absorption. In the $2p+3p$ case, the off-resonant contribution with $+$ is dominant, which implies that the slope of the phase is positive, in agreement with the results in Fig.~\ref{fig:ananum}~(b,e). We have found that increasing the photon energy of the pulse increases the steepness of the slope due to an increased relative contribution of the off-resonant terms. In the limit of only off-resonant contributions, the slope will become infinitely steep at the central frequency, as the phase for maximal absorption changes from $\pm\pi/2\rightarrow \mp\pi/2$.

\begin{figure}
\begin{center}
\includegraphics[width=0.33\textwidth]{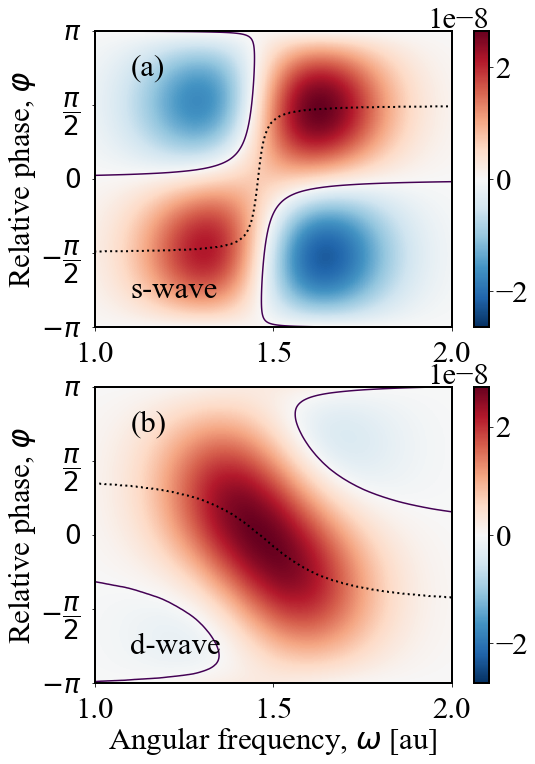}
\caption{Energy-resolved absorption of a hydrogen atom in the prepared superposition: 2p+3p, subject to an attosecond pulse with the intermediate angular momentum restricted to $s$-wave and $d-$wave in (a) and (b), respectively.}
\label{fig:angmon}
\end{center}
\end{figure}

\subsubsection{Role of angular momentum channels}
\label{sec:distang}
The absorption of a hydrogen atom in a $2p+3p$ superposition depends on the angular momentum channels $s$ and $d$, as shown in  \cref{fig: intro schematic}. The separated contributions to absorption and emission from the two partial waves are shown in \cref{fig:angmon}~(a) and (b) for the $s$ and $d$ channel, respectively. 
Similar to the $2s+3s$ case, the $d$ channel contribution from $2p+3p$ has a negative slope for maximal absorption. In contrast, the $s$ channel contribution has a positive slope with clear regions of emission that resemble the off-resonant $+$ contribution in Fig.~\ref{disentangle}~(d). Clearly, the $s$ channel dominates the total absorption and emission for the $2p+3p$ case. 

{\it The rule of slope:} We have found that off-resonant $+$ contributions dominate over $-$ contributions, when it is possible for an atom in a superposition to go to an intermediate state with lower energy. As an example, the $2p+3p$ case makes an off-resonant transition towards the $1s$ state, which means that the $+$ contribution will dominate and the slope will be positive.        
If there are no dipole-allowed intermediate states with lower energy, the off-resonant $-$ contribution will dominate and the slope will be negative.  
%
We have verified that the {\it rule of slope} is valid for general superpositions of two states with equal angular momentum ($\ell=\ell'$). Superpositions with higher angular momenta, such as the $3d+4d$ case, have a less dominant off-resonant $+$ contribution compared with the $2p+3p$ case. The reason for this is that the off-resonant $+$ contribution is more dominant if there is a dipole-allowed virtual state with lower energy and if the transition matrix element to this state is larger. Hence, if the energy difference is smaller or the transition is weaker, then the off-resonant $+$ contribution is less dominant.  

While the results shown in this subsection were computed for the hydrogen atom, we have verified that they exhibit the same behaviour for the helium atom in two-state superpositions: $1s^{-1}(2s+3s)$ and $1s^{-1}(2p+3p)$, using TDCIS theory \cite{greenman_implementation_2010}. In the next section, we study the more complex case of the neon atom, which has 6 electrons in the outermost $2p$ shell.

%

\begin{figure}
    \includegraphics[width = 0.5\textwidth]{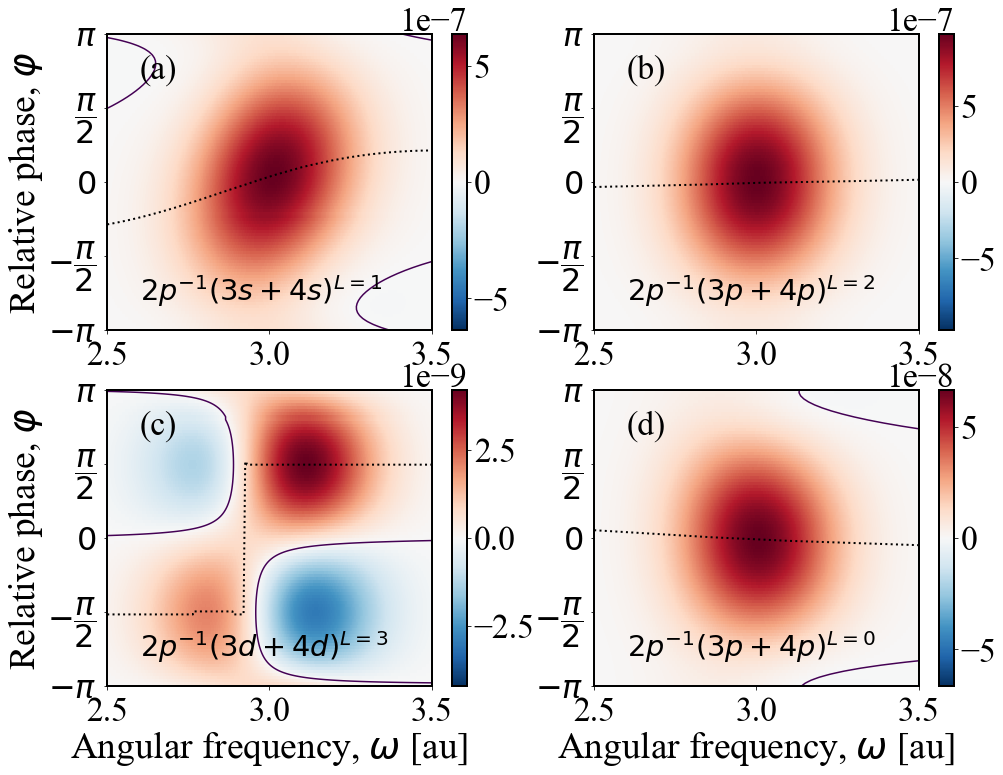}
    \caption{Energy-resolved absorption by a neon atom subject to an attosecond pulse in the superposition (a) $2p^{-1}(3s+4s)^{L=1}$, (b) $2p^{-1}(3p+4p)^{L=2}$, (c) $2p^{-1}(3d+4d)^{L=3}$ and (d) $2p^{-1}(3p+4p)^{L=0}$ with an estimated scale for RSP, $\varphi$. 
    The dotted black line shows the phase of maximal absorption (interpolation is used between discrete simulated values), which can be interpreted using the {\it rule of slope} (see main text). } 
    \label{fig:ne}
\end{figure}

\subsection{Neon atom} \label{Neatom}

The dynamics in neon atoms can be approximated by TDCIS theory \cite{greenman_implementation_2010}, provided that the role of double excited states are not essential for the physical process under consideration. While it is known that a detailed description of the ground state, containing double electron correlations, is essential for a quantitative description of one-photon ionization cross-sections of noble gas atoms \cite{amusia_atomic_1990,flugge_theory_1982}, the TDCIS theory provides a reasonable approximation for the neon atom. Rydberg states are found by diagonalizing the field-free problem including Coulomb interactions at the level of CIS to find eigenstates: $2p^{-1}n\ell^L$, where $L$ is the total angular momentum. The total magnetic quantum number is zero, $M=0$. Using the gerade ansatz for TDCIS \cite{pabst_impact_2012}, which provides a symmetry adapted basis for an atom excited by linearly polarized light, which are: $\ket{\Phi_{a,m=0}^{p,m=0}}$ and $\frac{1}{\sqrt{2}}(\ket{\Phi_{a,m=1}^{p,m=1}}+\ket{\Phi_{a,m=-1}^{p,m=-1}})$, where $m$ labels the magnetic quantum number of the hole (which equals that of the particle $m=m_a=m_p$). Here, we construct two-state superpositions of diagonalized states: $2p^{-1}(n\ell^L+n'\ell'^{L'})$ in the neon atom. In \cref{tab:nestates} the states used in the neon simulations are presented with the corresponding quantum numbers, symmetries and energies. Here $\abs{m^\mathrm{max}}$ is the most probable magnitude for the magnetic quantum number (probability in parentheses). The energies were validated with other computational methods and compared to NIST values, where a discrepancy is found due to the neglection of electron correlation in the CIS method.
\begin{table}[h]
    \centering
    \caption{Singly excited state energy levels of neon for the series $2s^{2}2p^{5}n'\ell'$ computed at the CIS level of theory.}
    \begin{tabular}{lccclcr}
       \hline\hline  \\
       Configuration  & $L$ & $\ell_{a}^\mathrm{max}$ & $\ell_{p}^\mathrm{max}$ & \multicolumn{1}{c}{ $|m^\mathrm{max}_{a,p}|$} & Sym. & \multicolumn{1}{c}{Level (eV) }    \\ \hline 
       $2s^2 2p^6   $ & 0   &     -        &       -      & \multicolumn{1}{c}{-} & g    & 0.0000  \\
       $2s^2 2p^5 3s$ & 1   & 1            & 0            & 0 (100\%)             & u    & 18.3625 \\
       $2s^2 2p^5 3p$ & 2   & 1            & 1            & 0 (67\%)              & g    & 20.1184 \\
       $2s^2 2p^5 3p$ & 0   & 1            & 1            & 1 (67\%)              & g    & 20.6010 \\
       $2s^2 2p^5 4s$ & 1   & 1            & 0            & 0 (100\%)             & u    & 21.2768 \\
       $2s^2 2p^5 3d$ & 1   & 1            & 2            & 1 (60\%)              & u    & 21.6172 \\
       $2s^2 2p^5 3d$ & 3   & 1            & 2            & 0 (60\%)              & u    & 21.6181 \\
       $2s^2 2p^5 4p$ & 2   & 1            & 1            & 0 (67\%)              & g    & 21.7613 \\
       $2s^2 2p^5 4p$ & 0   & 1            & 1            & 1 (67\%)              & g    & 21.9236 \\
       $2s^2 2p^5 5s$ & 1   & 1            & 0            & 0 (100\%)             & u    & 22.1503 \\
       $2s^2 2p^5 4d$ & 3   & 1            & 2            & 0 (60\%)              & u    & 22.2850 \\
       $2s^2 2p^5 4d$ & 1   & 1            & 2            & 1 (60\%)              & u    & 22.2862 \\
       \hline\hline
    \end{tabular}
    \label{tab:nestates}
\end{table}

Using Eq.~(\ref{FTTAS_ginv_VG}) we obtain the absorption and emission of a neon atom using the TDCIS approach with $2s$ and $2p$ as active orbitals, shown in \cref{fig:ne}. 
We investigate the prepared superpositions (a) $2p^{-1}(3s+4s)^{L=1}$, (b) $2p^{-1}(3p+4p)^{L=2}$, (c) $2p^{-1}(3d+4d)^{L=3}$ and (d) $2p^{-1}(3p+4p)^{L=0}$. We use a pulse with central frequency $\omega_0 = 3$ au to exclude resonant transitions to bound states, the other pulse parameters are the same as in \cref{Hatom}. As our method does not allow for assigning definite phases between the two initial states, we shift the phase to centre the absorption on zero phase. 
We clearly find that the \textit{rule of slope} describes the behaviour of the neon system, showing the applicability of the perturbative model on systems of higher complexity. We see a clear dominance of the off-resonant $+$ contribution for \cref{fig:ne}(a,c) due to the off-resonant transition to the $2p$ state. It is especially dominant for the $2p^{-1}(3d+4d)^{L=3}$ superposition presented in \cref{fig:ne}(c) as the transition between the $nd$ and $2p$ states is strong, due to the large overlap of the wave functions. 
However, for the $2p^{-1}(3p+4p)$ superpositions in \cref{fig:ne}(b,d) the off-resonant $+$ contribution is less dominant in accordance with the \textit{rule of slope} as the transition to the $2p$ hole is forbidden. In \cref{fig:ne}(d) the superposition $2p^{-1}(3p+4p)^{L=0}$ is constructed of mostly (67\%) $m=1$ orbitals, see \cref{tab:nestates}. As the $m$ quantum number is conserved, we therefore suppress the transition to the $s$ angular momentum channel, further limiting the number of dipole-allowed virtual states with energies below the prepared superposition (inhibiting transitions to $3s$). Hence, the off-resonant $-$ contribution is dominant in accordance with the \textit{rule of slope}. 
The effect of the $2s$ to $2p$ transition was determined by comparing with results where only the $2p$ orbital was active, finding that the effect was small.

\section{\label{conclusion}{Conclusion}}
In this work, we have presented a general gauge-invariant formulation of ATAS using the energy operator of Yang \cite{yang_gauge_1976}. This allowed us to unambiguously simulate absorption processes within a semi-classical description of light-matter interactions. In particular, we have considered the case of a hydrogen atom in a superposition state that is subjected to a weak attosecond pulse in the XUV regime that couples directly to the continuum.  We have constructed a model using perturbation theory that allows us to simulate the energy gain of atoms in both time and energy domains. It is found that the nature of the superposition, such as its quantum phases and angular momentum, determine the complex absorption process. Broad emission features are found in the energy domain with corresponding emissions in the time-domain being identified. Absorption processes are disentangled and it is shown that resonant contributions  are symmetric, while off-resonant contributions are anti-symmetric, with respect to the phase of the superposition. 
In more detail, the off-resonant contribution was shown to be dependent on the dipole-allowed virtual states and a \textit{rule of slope} was proposed to interpret the phase that maximizes the energy-resolved absorption of the attosecond pulse. Our model was validated by numerical simulations of TDSE for the case of the hydrogen atom. Simulations of helium and neon atoms were also performed, which indicated the applicability of our model to more complex atoms.   
Our model can be easily adapted to investigate weak absorption of light between bound states in atoms, but its strength lies at its proper treatment of continuum states. This may prove useful to study dynamics below, or across, the ionization threshold, in both time and energy. Further application of our model to study XUV absorption of laser-dressed atoms is a natural continuation of this work.     

\section{Acknowledgements} 

JMD acknowledges support from the Swedish Research Council: 2018-03845,
the Olle Engkvist Foundation: 194-0734 and the Knut and Alice Wallenberg
Foundation: 2017.0104 and 2019.0154.


\appendix

\section{ Matrix elements }\label{appendixC}

The bound-continuum and bound-bound matrix element of the momentum operator, $\hat p_z = - i \frac{d}{dz}$, are computed as 
\begin{widetext}
\begin{equation}
\label{Eq: TME}
\begin{split}
    \bra{f}\hat p_z\ket{j} 
    =&
    - i \bra{\ell'm'}\bra{R'_{\ell'}} \frac{d}{dz} \ket{\ell m}\ket{R_\ell}
    \\
    =& - i \delta_{\ell+1\ell'} \delta_{mm'}\sqrt{\frac{(\ell+m+1)(\ell-m+1)}{(2\ell+3)(2\ell+1)}} 
    \int_0^\infty r^2 R'_{\ell+1}(r) \frac{dR_{\ell}(r) }{dr} - \ell r R'_{\ell+1}(r) R_{\ell}(r) dr 
    \\
    &  - i \delta_{\ell-1\ell'} \delta_{mm'}\sqrt{\frac{(\ell+m)(\ell-m)}{(2\ell+1)(2\ell-1)}} 
    \int_0^\infty r^2 R'_{\ell-1}(r) \frac{dR_{\ell}(r)}{dr} + (\ell + 1) r R'_{\ell-1}(r) R_{\ell}(r) dr, 
\end{split}
\end{equation}
\end{widetext}
where the relation for the $z$ derivative of the product of the spherical harmonics and a generic $r$ dependent function $f(r)$ given in Eq. (A.37) in Ref. \cite{bethe_quantum_2013} has been used. $\ket{lm}$ are the spherical harmonics and $\ket{R_l}$ is either the radial wave function of the bound states in hydrogen, described by Eq. (3.17), or the continuum states described by energy normalized Coulomb waves, given by Eq. (4.23) in Ref. \cite{bethe_quantum_2013}. 

\section{ Resolving on a grid }\label{appendixB}

The energy domain absorption calculated using \cref{FTTAS_ginv_VG} with the momentum given by \cref{Eq: final momentum expectation value} can be written on the form 

\begin{equation}
    \frac{d \TAS }{d\omega} = C + K \delta(\omega\pm \Enjp) + Q/(\omega\pm \Enjp),
\end{equation}

\noindent
where the continuum states are contained in $C$, which we handle as constant in $\omega$ on small intervals. The bound state contribution is represented by $K$ for the resonant and $Q$ for the non-resonant contributions, the diverging parts are explicitly given. In order to handle the diverging elements, we represent the absorption on the numerical grid as 
\begin{equation}
\begin{split}
    \frac{\Delta \TAS }{\Delta \omega} = &
    \frac{1}{\Delta\omega} 
    \int_\omega^{\omega + \Delta\omega} d\omega \frac{d \TAS }{d\omega}
    \\
    =& C + \frac{K}{\Delta \omega} 
    \delta_{n,(n_{j'}^{-2} \pm 2\omega)^{-\frac{1}{2}}} 
    \\
    &+
    \frac{Q}{\Delta\omega} \int_{\omega}^{\omega+\Delta\omega}  
    d\omega' \frac{\text{p.v.}}{\omega' \pm \Enjp},
\end{split}
\end{equation}
where $\Delta \omega$ is the resolution of the grid, and we treat the singularity in the integral as a principal value. 


\bibliography{paper}

\end{document}